\documentclass[conference]{IEEEtran}
\IEEEoverridecommandlockouts
\usepackage{cite}
\usepackage{amsmath,amssymb,amsfonts}
\usepackage{algorithmic}
\usepackage{graphicx}
\usepackage{textcomp}
\usepackage{xcolor}
\def\BibTeX{{\rm B\kern-.05em{\sc i\kern-.025em b}\kern-.08em
    T\kern-.1667em\lower.7ex\hbox{E}\kern-.125emX}}
\begin{document}

\title{Overcoming Inverse-square Law of Gravitation and Luminosity for Interstellar Hyperspace Navigation by Celestial Objects}

\author{\IEEEauthorblockN{Nghi C. Nguyen}
\IEEEauthorblockA{\textit{HSSS Science} \\
Virginia, USA \\
nghi@hsss.science\\
https://orcid.org/0000-0003-3913-1828}
}

\maketitle

\begin{abstract}
%This document is a model and instructions for \LaTeX.
%This and the IEEEtran.cls file define the components of your paper [title, text, heads, etc.]. *CRITICAL: Do Not Use Symbols, Special Characters, Footnotes, or Math in Paper Title or Abstract.
Interstellar travel with hyperdrive or warp drive technologies presents navigating challenges when traveling speed approaches or exceeds the speed of light. Conventional methods such as inertia, magnetic, controlled gyros, visual or radar navigation, all have their limit at near-light speed due to inverse-square laws for gravitation, luminosity, and isotropic radiator. Near-field gravitational force of Alcubierre Warp Drive and Hyperdrive makes inertia, magnetic, and controlled gyros ineffective within the hyperbolic curvature of spacetime's strong gravitational and magnetic alignment. 

In this paper, we propose novel methods to increase total luminosity and decrease error propagation during subluminal or luminal interstellar transits by navigation with star maps. We demonstrate with theoretical and simulation models in first order approximation results that are correlated to eccentric and concentric of light bands, ebbing and ascending of light streaks, and centric of celestial distance. Finally our models show that both near-field and far-field celestial objects enhance error correction and navigation vectors even when the traveling speed is approaching the speed-of-light. We demonstrate that our method of varying a spaceship's primary axis rotational rate and angle of approach are superior than traveling at a straight line, especially when the traveling speed is approaching or exceeding the speed of light.
\end{abstract}


\section{Background}
Navigating in hyperspace presents many challenges. Spacetime distortion at luminal or super-luminal speed by Special Relativity and near-field effects of gravitational tidal forces by General Relativity make conventional navigating methods error-prone \cite{REF_hawking1973large}. A small degree of error during hyperspace navigation would cause either parsec distance misses or worse, a spaceship coming out of hyperspace right on top of a planet or a star.

Conventional methods such as inertia, magnetic, controlled gyros, visual or radar navigation, all have their limit at near-light speed due to inverse-square laws for gravitation (Newton’s laws), luminosity (light intensity), and isotropic radiator (radio \& radar). Near-field gravitational force of Alcubierre Warp Drive and Hyperdrive makes inertia, magnetic, and controlled gyros ineffective within hyperbolic curvature of spacetime's strong gravitational and magnetic alignment \cite{REF_EinsteinSingularities}. 

Navigation by star maps of celestial is still the ideal method in interstellar travel because the relative incident angle of arriving star lights faraway do not change significantly relative to traveling distance. In this paper, we propose a novel method to overcome the inverse-square law of gravitation and luminosity. When we apply both rotational rates of approach, major angle $\alpha$ and minor $\omega$, in line with traveling vector, star map quality increases significantly due to total illumination over larger area. Uniquely resulting star lights aggregate into various navigational bands or streaks that further enhance error correction for controlling the spacecraft. Monitoring spacetime dilation of light streaks over time and correlating with wavelength spectral shift act as control feedback to the Guidance, Navigation and Control (GNC) system of a spacecraft. 

\section{In-system Transit versus Interstellar Navigation}
In-system or inner-system transit present a different set of navigation and avoidance strategies than inter-system or interstellar traveling. Within a single or binary star system, its stars act as primary light sources as well as illuminating any large celestial objects orbiting this system; their near-field illumination dominate the quality and accuracy of star maps. For within a star system, there are a sufficient quantity of illuminated celestial objects with known orbiting planes to give a precise location of a spacecraft relative to its star.

Secondly, it is not wised to engage warp-drive or hyperdrive for inner-system traveling between in-system planets for long duration. Traveling at luminal speed would be very short in relative distance to those of a star system, as well increasing the risk of collision with other in-system objects. A preferred method is to engage short jumps for precise traveling; this method eliminates the need for enduring navigation using warp-drive or hyperdrive.

Star systems are sparsely populated in our universe, their sizes are minuscule to large empty space and black void separated by light-year distances. Interstellar travel with warp-drive and hyperdrive between the nearest systems would take months or perhaps years. In luminal or super-luminal--also known as faster-than-light (FTL)--mode traveling, all light sources are far-field and subject to inverse-square law of luminosity reducing their quality, as well as spacetime distortions induced by Special Relativity \cite{REF_Wormsholes}. Most critically, warp-drive and hyperdrive generate near-field gravitational tidal forces that are dominating spaceship's perceptibility of arriving star lights under inverse-square law of gravitation by General Relativity's spacetime curvature. This paper applies to this type of interstellar traveling.

\section{Effects of Near-field Gravitational Sources under General Relativity}
Both Alcubierre's warp drive \cite{REF_Alcubierre} and the author's hyperdrive \cite{REF_NguyenKey} require near-field sources of gravitational matter and anti-matter. This section details the near-field effects of strong gravitational sources under General Relativity \cite{REF_NguyenNearField}, and proposes various techniques to enhance navigation by celestial lights during hyperspace traveling mode. 

\subsection{Usable Field-of-View}
\begin{figure}[htbp]
\centerline{\includegraphics[width=3.5in]{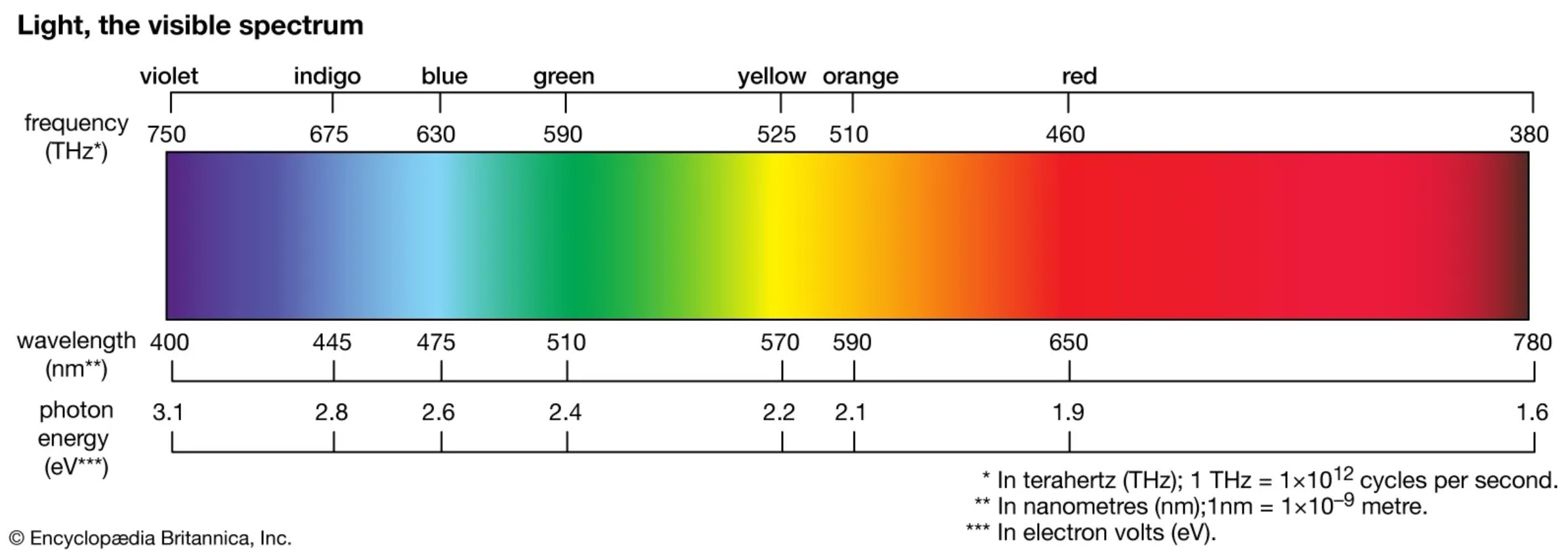}}
\caption{Visible Spectrum of Light \cite{IMG_LightSpectrum}.}
\label{VisibleLight}
\end{figure}

Interstellar navigation by celestial objects require an unobstructed view to the star fields in the frame of the traveling spacecraft, defined as usable field-of-view (FOV). There are many types of instruments to collect star lights, from wide angle camera systems and crude devices such as coarse sun sensor to precision focal systems such as star trackers, that can detect spectral bands from visible to non-visible such as near-infrared and x-ray. In this paper, we use visible light band as the reference spectrum with wavelength between 400nm to 780nm, ``Fig.~\ref{VisibleLight}".

For transit in-system, a spaceship with an array of light receivers strategically positioned around its hull would have unfettered access to a 360 degree FOV, with sun shields to block out powerful rays from the in-system primary star. Interstellar travel with conventional propulsion technologies rarely needs sun shield deployment due to the majority of transit time being in the void or empty black space between star systems. However, warp drive and hyperdrive generate massive gravitational sources in close-distance to the spaceship, and strong gravitational tidal near-field forces make gravitational lensing that limit the usable FOV \cite{REF_NguyenKey}.

\subsection{Alcubierre's warp drive}
\begin{figure}[htbp]
\centerline{\includegraphics[width=2.5in]{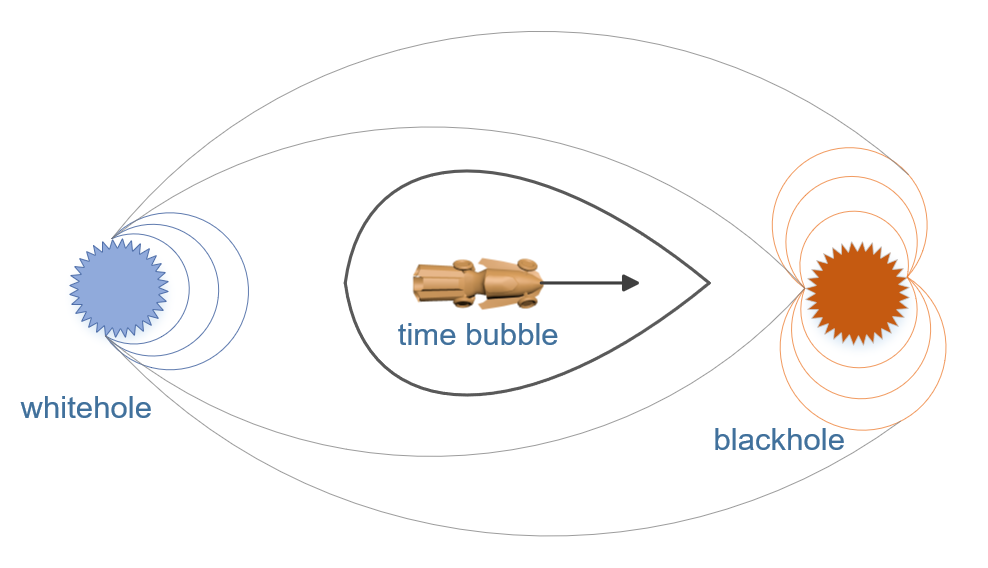}}
\caption{Conceptual Alcubierre's Warp Drive.}
\label{AlcubierreDrive}
\end{figure}

Alcubierre's warp drives require equal and opposite gravitational fields generated by matter (blackhole) and anti-matter (whitehole) as shown in ``Fig.~\ref{AlcubierreDrive}". A time-stable bubble envelopes the spaceship, protecting it from time dilation due to Special Relativity effect of FTL traveling speed. The aft whitehole is generated by concentrated anti-matter to produce space expansion behind the ship pushing it; at the same time  the blackhole by concentrated matter produces a contraction of space pulling the spaceship forward. This push-pull effect from space expansion and opposite contraction requires close proximity of the whitehole and blackhole to the spaceship in equal distance. It requires a dedicate balancing control to keep blackhole and whitehole apart, as anti-matter and matter annihilate themselves and the spaceship in contact.

Within the time bubble, spaceship is buffered by strong gravitational fields that subject the front and rear views to strong and weak gravitational lensing. Gravitational lensing makes far-field celestial objects appear in different locations than their true coordinates. Error of near-field General Relativity increases proportionally with gravitational lensing angle, compounding position dilution of precision for navigation.

\begin{figure}[htbp]
\centerline{\includegraphics[width=2.5in]{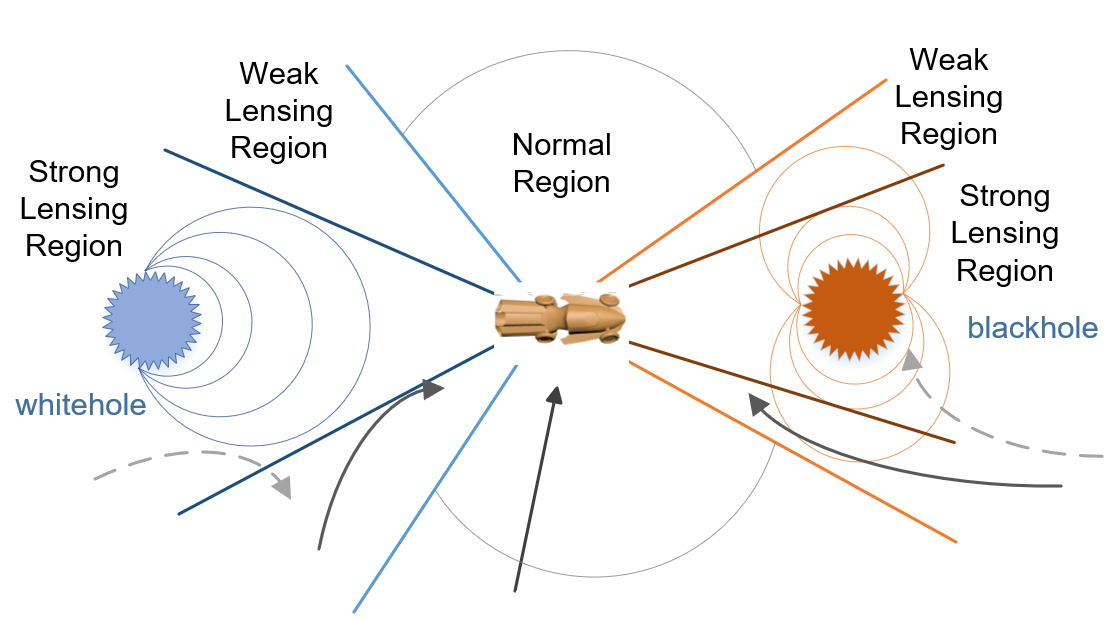}}
\caption{Alcubierre Field-of-View Regions.}
\label{AlcubierreFOV}
\end{figure}

 In the case of Alcubierre's warp drive, the gravitational lensing regions behind the spaceship are larger than in the forward lensing region; aft expansion of space pushes arriving lights away from the spaceship, as opposed to pulling into for the case of forward space contraction ``Fig.~\ref{AlcubierreFOV}". Ideal FOV to triangulate traveling spaceship with the most precision of position are light arriving from the normal region without gravitational lensing. Therefore, FOV of normal region by Alcubierre spacecraft during hyperspace would be limited to the radial axis but not the axis transverse to the direction of travel.

\begin{figure}[htbp]
\centerline{\includegraphics[width=2in]{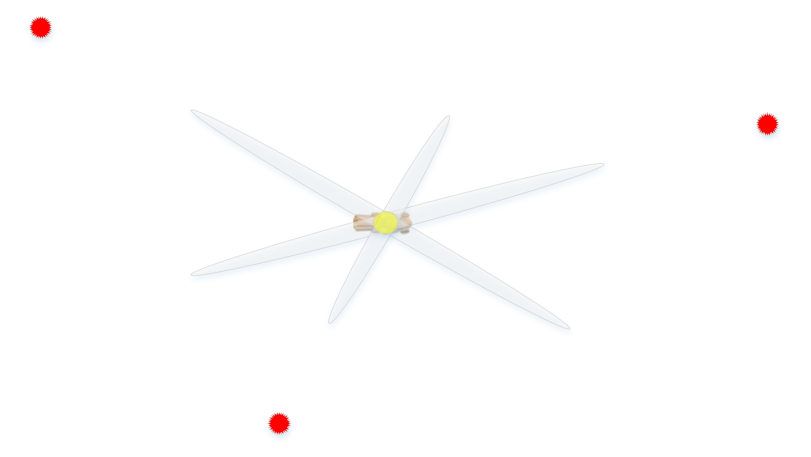}}
\caption{Ideal Separation in 2D FOV.}
\label{IdealFOV}
\end{figure}

Computing Euclidean coordinates of a traveling spacecraft in 3-dimensional (3D) space using known star map for navigation, requires a minimum of three reference stars in FOV. Ideally the three reference stars are noncoplanar, maximally separated and orthogonal in their 3D planes, which yield the least circular error propagation, ``Fig.~\ref{IdealFOV}". 

\begin{figure}[htbp]
\centerline{\includegraphics[width=2in]{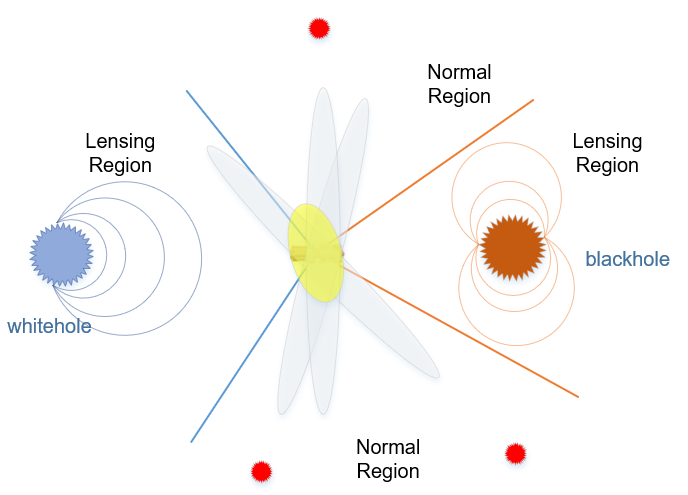}}
\caption{Alcubierre Radial Error due to Limited FOV.}
\label{AlcubierreError}
\end{figure}

For Alcubierre spacecraft, its limited FOV forward and backward in the transverse axis leads to the most error propagation in radial dimensions orthogonal to the transverse axis, ``Fig.~\ref{AlcubierreError}". The circular error or position dilution grows proportionally with gravitational field strength, distance to the gravitational field, and relative traveling speed.

\subsection{Hyperdrive with $\theta$-offset}
\begin{figure}[htbp]
\centerline{\includegraphics[width=2in]{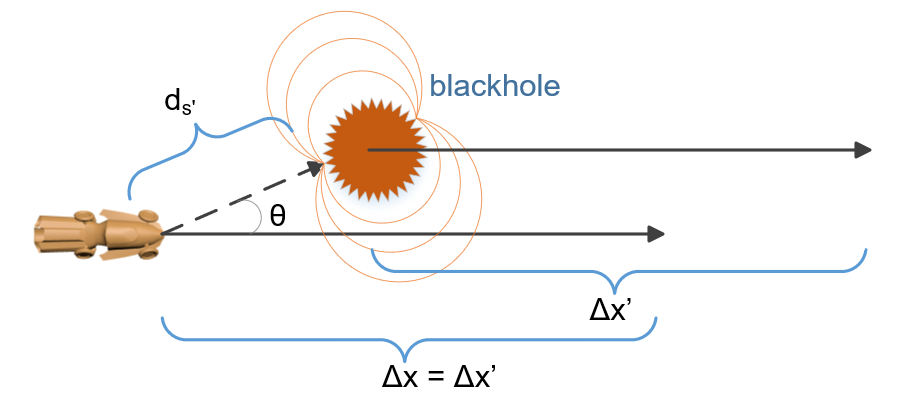}}
\caption{Conceptual of Hyperdrive Spacecraft.}
\label{SpaceshipHyperdrive}
\end{figure}

The author proposed a concept of hyperdrive spaceship that does not require exotic anti-matter \cite{REF_NguyenKey}. Without aft space expansion, the spacecraft is required to keep a fixed distance $d_{s'}$ to forward blackhole with an angle $\theta$ from the transverse axis of travel, ``Fig.~\ref{SpaceshipHyperdrive}". As spacecraft continuously falls into forward blackhole and regenerative blackhole moves away keeping their relative distance $d_{s'}$ constant, a time-stable bubble envelops both the spacecraft and the blackhole as observed from the third person frame.

\begin{figure}[htbp]
\centerline{\includegraphics[width=2in]{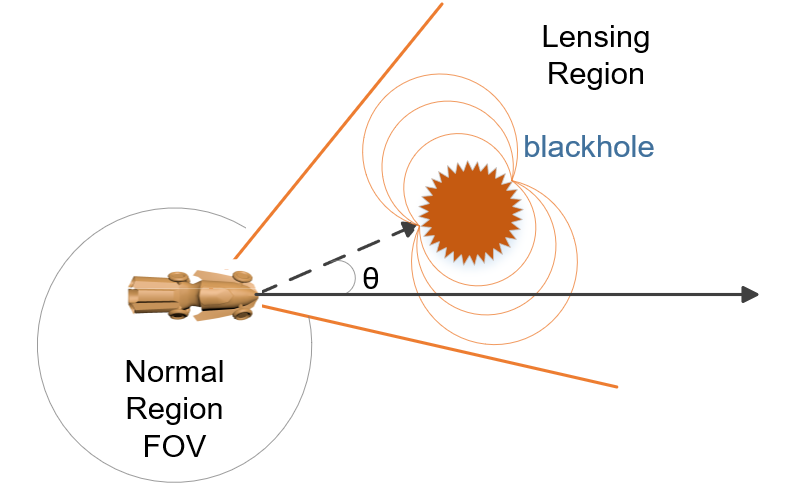}}
\caption{Hyperdrive Spacecraft FOV.}
\label{HyperdriveFOV}
\end{figure}

The hyperdrive spaceship has full access to 180 degree aft region; its FOV encompasses larger polar angles than Alcubierre's. With larger FOV, circular error propagation is decreased in both the radial and transverse axis, ``Fig.~\ref{HyperdriveFOV}".  

The spaceship needs to trade off between $\theta$ angle for operating safe margin, relative traveling speed, and FOV. Their relationship is such that larger $\theta$ improves safety for spaceship to operate in hyperspace but at the cost of slower linear relative speed and reduced FOV.

\subsection{Hyperdrive with $\theta$-offset and $\omega$-rotation}
\begin{figure}[htbp]
\centerline{\includegraphics[width=2in]{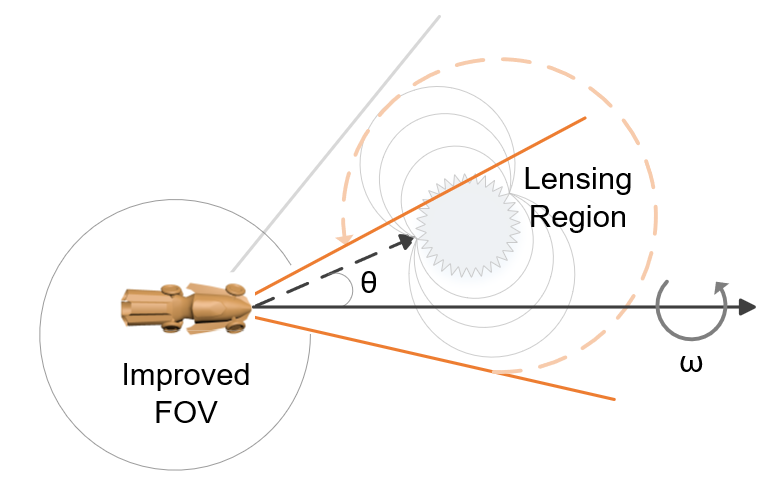}}
\caption{Improving FOV cone with $\omega$-rotation.}
\label{HyperdriveFOVimrpoved}
\end{figure}

Hyperdrive ship can effectively increases its FOV by adding an $\omega$ rotation around the transverse axis in line with the traveling direction, ``Fig.~\ref{HyperdriveFOVimrpoved}". Due to small angle gained by $\theta$-offset from traveling vector, rotating FOV around its transverse axis yields smaller angular cone of lensing region polar opposite to $\theta$ traverse angle by the rate of $\omega$ over time.

\subsection{2D View Projection by Horizontal Coordinate System}
\begin{figure}[htbp]
\centerline{\includegraphics[width=3in]{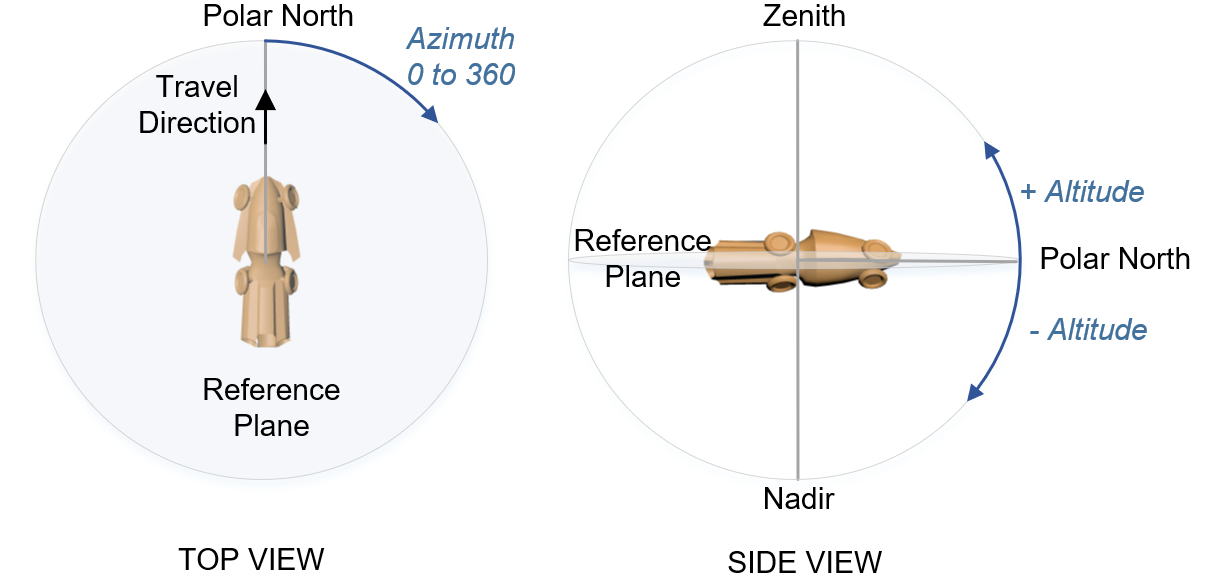}}
\caption{Horizontal Coordinate System.}
\label{HorizontalRef}
\end{figure}

For flight instrumentation, it is necessary to use Horizontal Coordinate System for celestial navigation. Take a polar coordinate system: Polar North is travel vector, Zenith is above the spaceship's observer bridge and Nadir below, with bridge floor is the reference plane or horizontal plane, ``Fig.~\ref{HorizontalRef}". 

\begin{figure}[htbp]
\centerline{\includegraphics[width=1.5in]{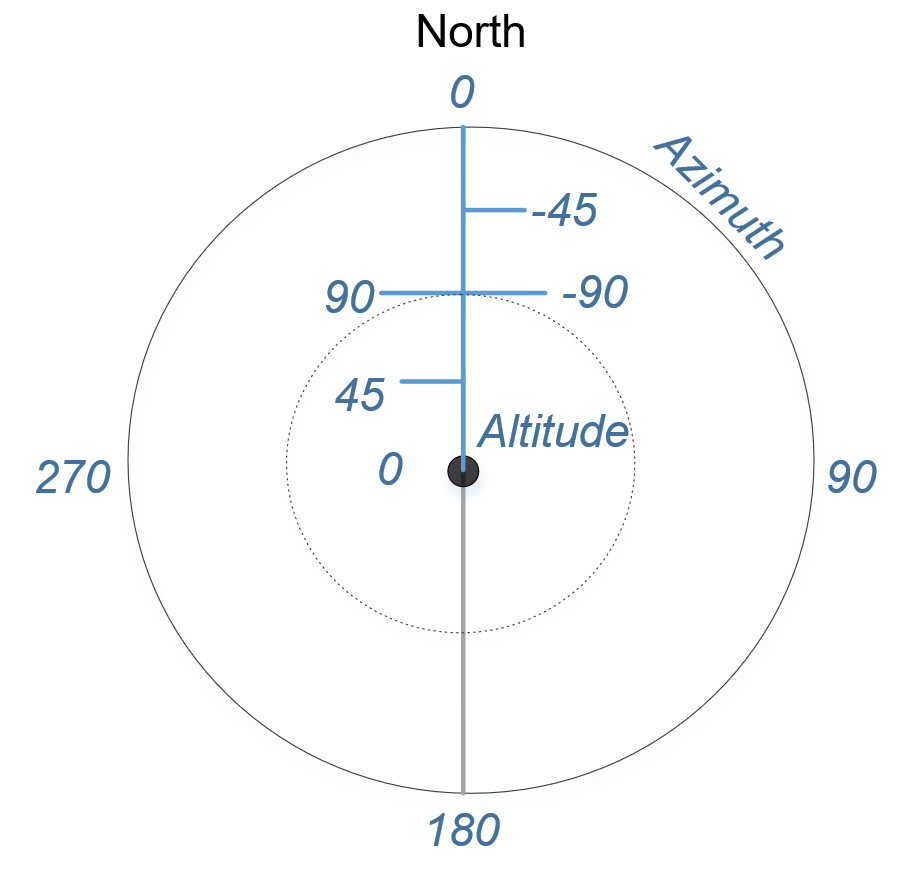}}
\caption{HUD Flight Instrument of Horizontal Coordinate System.}
\label{HorizontalProjection}
\end{figure}

Using the spaceship's observer bridge as the reference point, Azimuth is the horizontal angle of the star from Polar North projected onto the horizontal plane. Attitude, also known as elevation, is the angle between the star to horizon plane. For head-up display (HUD) of flight instruments, a 3D projection onto 2D polar coordinates, ``Fig.~\ref{HorizontalProjection}", carry all the information with 360 degree FOV around the spaceship. The 2D HUD contains tracking objects in degrees by Cartesian coordinate as (azimuth, altitude) translated to Polar coordinate as $(\rho, \phi)$, for $0^{\circ} \leq azimuth \leq 360^{\circ}$ and  $-90^{\circ} \leq altitude \leq 90^{\circ}$, and their relationship:

\begin{equation}\label{EQ_PolarConversion}
\begin{aligned}
azimuth = & \rho * cos (\phi) \\
altitude = & \rho * sin (\phi)
\end{aligned}
\end{equation}

\begin{figure}[htbp]
\centerline{\includegraphics[width=1.5in]{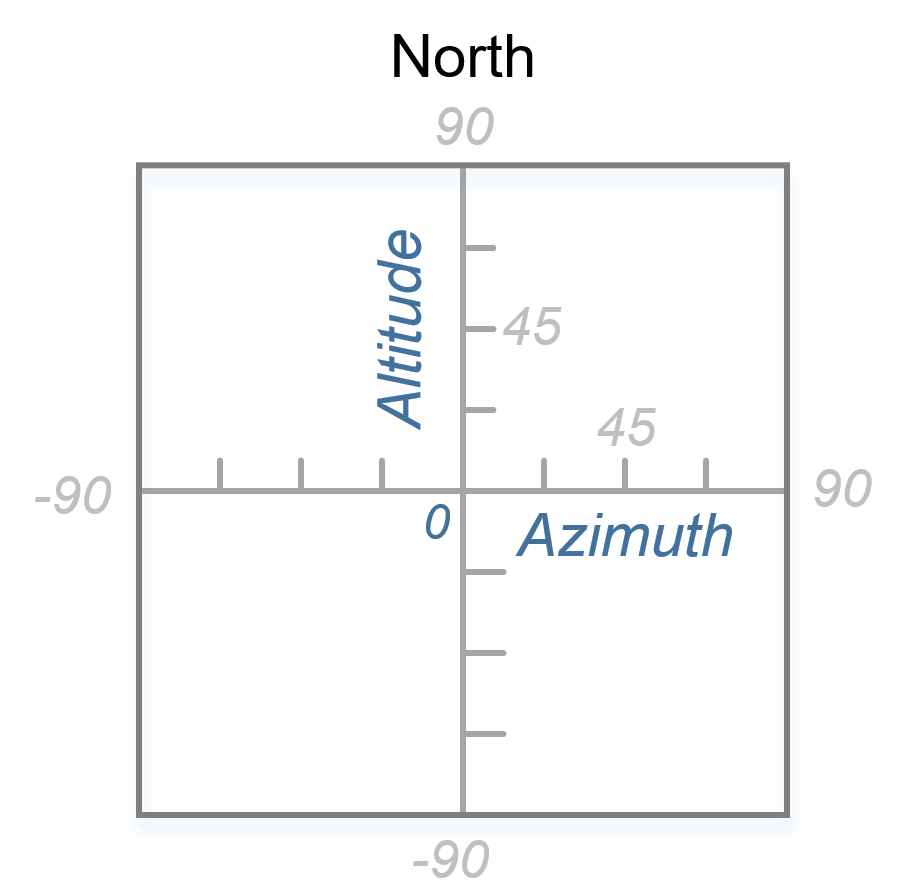}}
\caption{FOV in 2D Cartesian Form.}
\label{2DFOV}
\end{figure}

To calculate error propagation of reference star map during flight, it is necessary to project 3D Euclidean coordinates with the spaceship as the reference frame onto a 2D Cartesian plane orthogonal to the transverse axis of travel, ``Fig.~\ref{2DFOV}". This 2D Cartesian plane limits FOV display to half of the hemisphere, to either forward and aft sections. It is easier for human to perceive parallel projection of 2D Cartesian view than instrument HUD view, hawking to star field from the view point on Earth.

These 2D Cartesian coordinates are expressed in degrees in (azimuth, altitude) converted from polar form $(\rho, \phi)$, limited to $-90^{\circ} \leq azimuth, altitude \leq 90^{\circ}$, or in radians, $-\pi/2 \leq azimuth, altitude \leq \pi/2$.

\subsection{Effective Gain in Error Distribution Lapse over Time}
\begin{figure}[htbp]
\centerline{\includegraphics[width=3in]{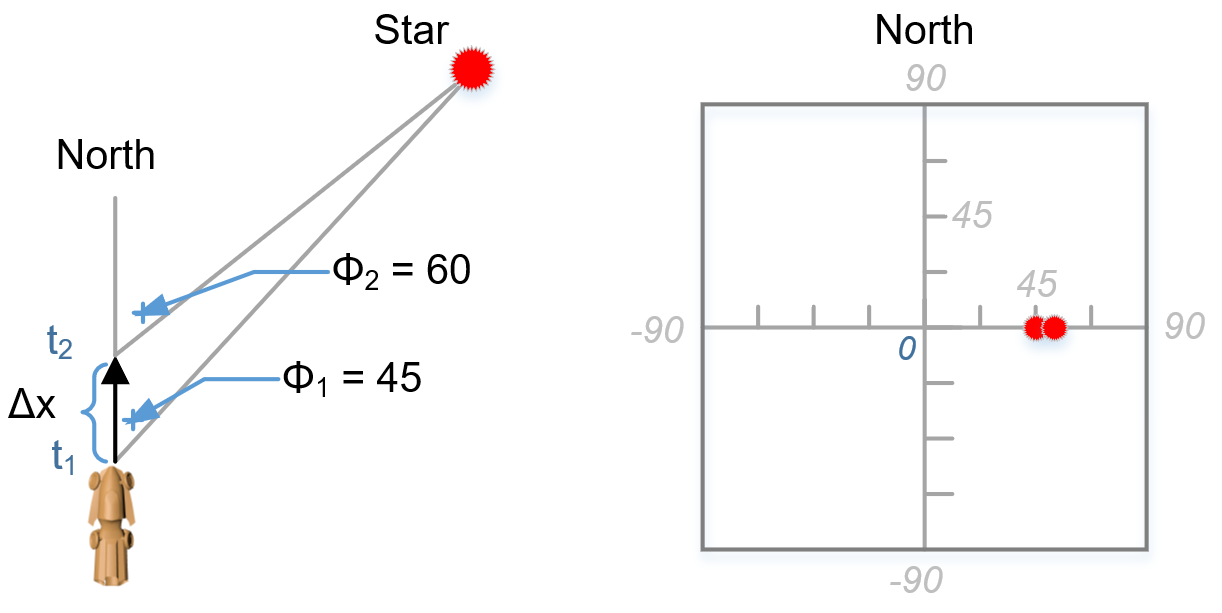}}
\caption{Example of Star Transiting in 2D FOV.}
\label{StarFOV}
\end{figure}

Take a reference star with a nominal distance $d$=10 parsecs, azimuth = +45$^{\circ}$, and altitude = 0$^{\circ}$ in horizontal coordinate at time $t_1$; after traveling a distance $\Delta x$ at time $t_2$, this star's new position has azimuth = +60$^{\circ}$, and altitude = 0$^{\circ}$. Without accounting for relativistic shift of luminosity, FOV changes from $t_1$ to $t_2$ are shown in ``Fig.~\ref{StarFOV}".

\begin{figure}[htbp]
\centerline{\includegraphics[width=3in]{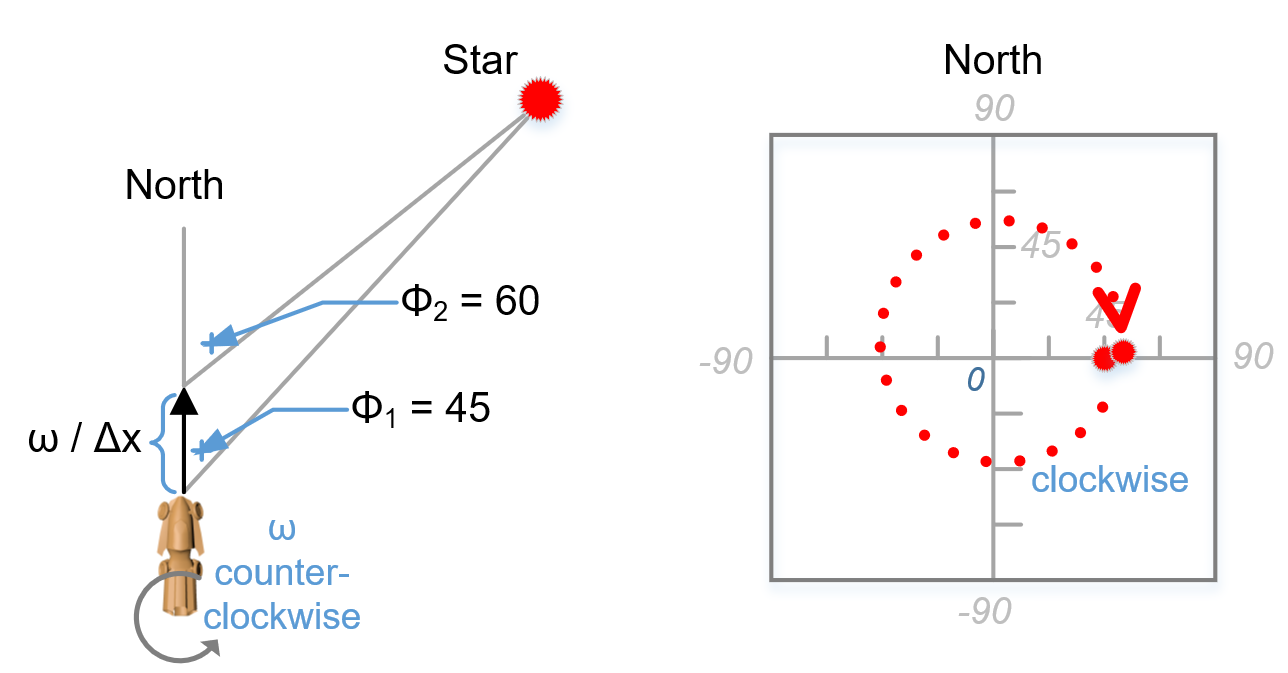}}
\caption{Example of Star Transiting with $\omega$ Rotation.}
\label{StarFOVrotate}
\end{figure}

Adding a full 360$^{\circ}$ clockwise $\omega$ rotation around the transverse axis in line with the traveling direction between $t_1$ to $t_2$, 2D FOV from the reference of the spacecraft is shown in Fig.~\ref{StarFOVrotate}, with the star transiting in clockwise direction, by converting polar coordinates to Cartesian using trigonometric functions of Eq. \ref{EQ_PolarConversion}.

Perceived reference star luminosity from the spacecraft is apparent magnitude $m$ that classified the level of brightness of stars, for which Vega is the standard reference star with $m=0$ as viewed from Earth. The effective error propagation distribution $\epsilon$ of a star tracking through 2D Cartesian plot is its apparent magnitude $m$ collected over distance $\Delta x$ over time $\Delta t$:

\begin{equation}\label{EQ_error}
\epsilon = \Sigma_{\Delta t} (m) \cdot \omega \cdot \frac{\Delta x}{\Delta t}
\end{equation}

From ``Fig.~\ref{StarFOVrotate}", the effective error propagation distribution gained by using 2D FOV of celestial reference over time lapse $\Delta t$ by a full rotation $\omega$ over travel distance $\Delta x$, is the circumference distance by averaging:

\begin{equation}\label{EQ_ErrorDistance}
\Delta x = 2\pi \cdot  \rho_{average} = 0.583 \pi^2
\end{equation}

which $\rho$ is polar radius in radian unit as:

\begin{equation}\label{EQ_ErrorRadius}
\rho_{average} = \frac{(45+60)}{2} \cdot \frac{\pi}{180} = 0.292 \pi
\end{equation}

Compared with no rotation from ``Fig.~\ref{StarFOV}", error propagation distribution over distance time lapse for $\Delta x_{\omega=0}$ is:

\begin{equation}\label{EQ_ErrorRadiusNoRotation}
\Delta x_{\omega=0} = (60-45) \cdot  \frac{\pi}{180}  = 0.083 \pi
\end{equation}

As $m$ is the same apparent magnitude perceived from the spacecraft FOV for both cases, $\omega=0$ and $\omega=2\pi$, thus the effective gain of one rotation over the same traveling distance $\Delta x$ over time $\Delta t$ is:

\begin{equation}\label{EQ_ErrorGain}
Gain_{\epsilon} = \frac {0.583\pi^2} {0.083 \pi} \approx 7\pi \approx 22
\end{equation}

Error propagation is further degraded when accounting for inverse-square law of luminosity and by Special Relativity at luminal traveling speed, which leads to next section.

\section{Effects of Relativistic Doppler under Special Relativity}

\subsection{Wien's Displacement Law}
Stefan-Boltzmann Law defines total radiant power emitted from a surface proportional to the fourth power of its absolute temperature, with $E$ being the radiant heat energy emitted, $T$ the absolute temperature (K), and $\sigma$ Stefan-Boltzmann constant of $5.670374419 \cdot 10^{- 8} (W/m^2/K^4)$ as:

\begin{equation}\label{EQ_StefanBoltzmann}
E = \sigma T^4
\end{equation}

German physicist Wilhelm Wien found the inverse relationship between the temperature of a blackbody and the wavelength at which it emits the most light. Wien's Displacement Law states that the spectral radiance of black-body radiation per unit wavelength reach a maximum at a certain wavelength $\lambda_m$ by an inverse proportion to temperature $T$, with $b$ Wien's displacement constant of $2.897771955 \cdot 10^{- 3} (mK)$ as:

\begin{equation}\label{EQ_Wien}
\lambda_m = b / T
\end{equation}

\begin{figure}[htbp]
\centerline{\includegraphics[width=3.25in]{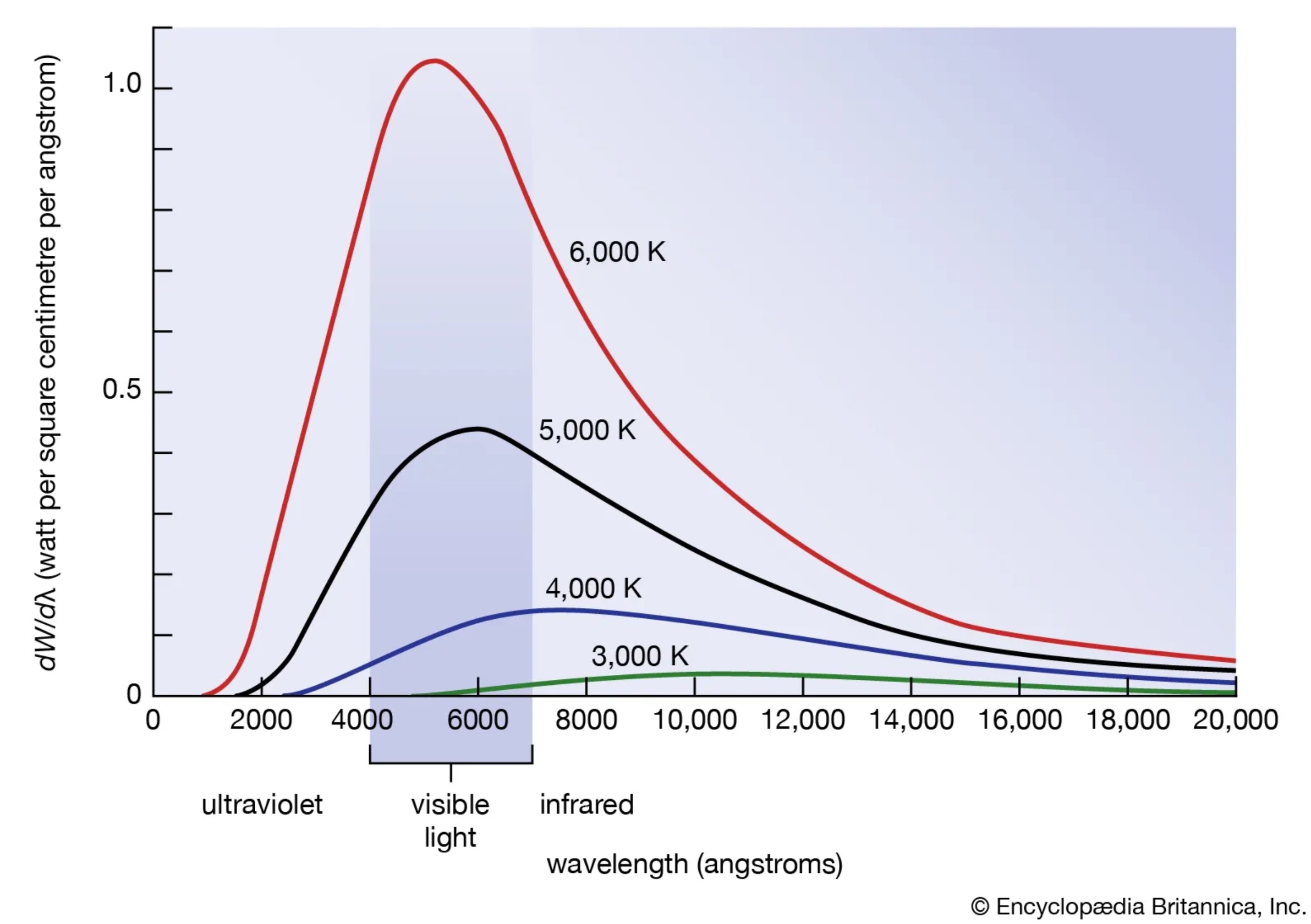}}
\caption{Blackbody Radiation Chart of Wien's Displacement Law\cite{IMG_WienLaw}.}
\label{WienLaw}
\end{figure}

Thus the inverse relationship between temperature and peak wavelength of blackbody radiation is shown as ``Fig.~\ref{WienLaw}". Each visible star would have a unique spectral and peak wavelength by Stefan-Boltzmann and Wien's Laws; their varied mass, size, and age invariably dictate their surface temperature, which correlate with their absolute spectral magnitude or intrinsic luminosity. The star's unique spectral magnitudes or spectral fingerprints can aid our navigation by referencing their spectral peak $\lambda_m$ and relative luminosity $m$. However, the spectral peaks are subjected to the relativistic Doppler effect proportionally to traveling speed $\beta=v/c$, and relative luminosity $m$ subjected by inverse squared distance.

\subsection{Hertzsprung-Russell Diagrams}
\begin{figure}[htbp]
\centerline{\includegraphics[width=3in]{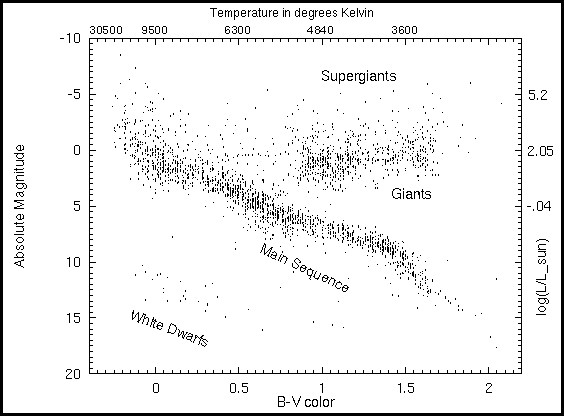}}
\caption{Hertzsprung-Russell Diagram \cite{IMG_NickStrobelHR}.}
\label{HRdiagram}
\end{figure}

The relationships of celestial reference in distance, luminosity,  temperature, and spectral can be captured in a Hertzsprung-Russell diagram, or H-R diagram. The diagram is named after astronomers, Ejnar Hertzsprung and Henry Norris Russell who independently developed the plot in early 1912 that show correlation between the majority of stars lie along a diagonal band called main sequence, ``Fig.~\ref{HRdiagram}".

\begin{figure}[htbp]
\centerline{\includegraphics[width=3in]{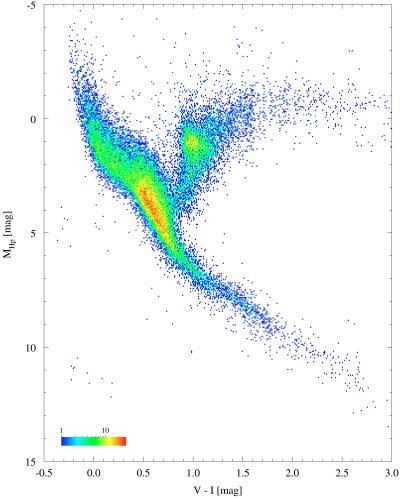}}
\caption{Actual H-R Diagram Based on Hipparcos Data \cite{IMG_HRdiagram}.}
\label{ESAhrdiagram}
\end{figure}

The H-R diagram is useful for comparative charting of stars by their characters, absolute magnitude (intrinsic luminosity) and spectral class based on surface temperature. Astronomer Annie Cannon while at Harvard classified star spectral sequence by their temperature, $O B A F G K M$. This spectral sequence is known as the Harvard classification, with $O$ stars being the hottest with temperatures of around 40,000 Kelvin (K) and $M$ stars the coolest at around 2,500K; our Sun is a G2 type with and a surface temperature of around 6,000K, ``Fig.~\ref{ESAhrdiagram}".   

\subsection{Inverse-square Law of Luminosity}

By Conservation of Energy, the total energy $P$ of a point source stellar irradiance emanated over a spherical surface is constant as:

\begin{equation}\label{EQ_TotalEnergy}
P = |I| \cdot A_{surface} = |I| \cdot 4\pi d^2
\end{equation}
where $d$ is the distance from the point source, and $|I|$ is the intensity function at the surface. This gives intensity of point source at distance $d$ perceived at the spaceship as:

\begin{equation}\label{EQ_Intensity}
|I| = \frac{P}{4\pi d^2} 
\end{equation}

Apparent energy $P_m$ perceived at the receiver from a distance $d$ is the apparent magnitude $m$, and apparent radius $r_m$ is the radius of apparent star light. Thus the 2D projected Cartesian area $a=\pi r_m^2$ orthogonal to starlight vector perceived by the spaceship's receiver is the total apparent energy, a function of intensity $|I|$ and surface area $a$ as:

\begin{equation}\label{EQ_ReferenceMagnitude}
P_m = |I| \cdot a = \frac{P}{4\pi d^2} \cdot \pi r_m^2 = \frac{r_m^2}{4d^2} \cdot P
\end{equation}

Thus, we derive the ratio of $m$ and $M$ as a function of distance $d$ as:

\begin{equation}\label{EQ_MagnitudeRatio}
\frac{m}{M} = \frac{P_m}{P}= \frac{r_m^2}{4d^2}
\end{equation}

or

\begin{equation}\label{EQ_MagnitudeDistance}
m = \frac{r_m^2}{4d^2} \cdot M
\end{equation}
which proves that apparent magnitude has an inverse-square distance relationship to absolute magnitude that agrees with the Inverse-square Law of Luminosity. 

Therefore, a referenced star can be projected onto the 2D FOV from 3D polar coordinates as $(m,\theta)$ where $m$ is a function $F(r_m,d,M)$, $d$ is spaceship-star distance, $M$ is star's absolute magnitude, and $\theta$ is a pair projected 2D Cartesian field of ($\theta_{azimuth}, \theta_{altitude}$); apparent radius $r_m$ is a function of the receiver's focal length and sensitivity in filter $F_x$ band analog to digital conversion of measured apparent radiant area (Eq. \ref{EQ_ApparentMagnitude}). These referenced stars can be projected onto 2D Cartesian FOV Plot of ``Fig.~\ref{2DFOV}", with apparent magnitude $m$ represented as area circle $r_m$ and color coded for intensity as a function of absolute magnitude $M$ adjusted for inverse-square distance $d$. 

From Eq. \ref{EQ_ErrorGain}, we can now apply distance effect to Error Propagation Distribution $Gain_{\epsilon}$ using Eq. \ref{EQ_MagnitudeRatio} of distance $d^2$ by trigonometric functions of $\Delta x$ and $\theta$. Given reference stars are significantly far compared to travel distance, $d \gg \Delta x $, $\Delta d$ distance has minimal effect to $Gain_{\epsilon}$ function when angular rotation $\omega$ is applied.

\subsection{Exploiting Stellar Spectra Shift}

Referenced star's relative luminosity or apparent magnitude $m$ received by spectral receivers is normalized by reverse logarithmic scale where $F_x$ is the observed irradiance using spectral filter $x$ and $F_{x,0}$ the calibrated reference flux of the receiver within spectral band $x$

\begin{equation}\label{EQ_ApparentMagnitude}
m_x = -5 log_{100} \frac{F_x}{F_{x,0}}
\end{equation}

Stellar bolometric magnitude $M_E$ derived from Eq. \ref{EQ_StefanBoltzmann}, is the total of all radiation emitted by a star at all wavelengths. It is inefficient to use bolometric magnitude for navigation due to receiver's limited bandwidth, a trade off between filter $x$ thermal noise, focal length, and aperture sensitivity. Thus we limit stellar spectra $M_{visual}$ to the visible band, ``Fig.~\ref{VisibleLight}".

From ``Fig.~\ref{ESAhrdiagram}", the majority of stars (90\%) have unique stellar spectral fingerprints with peak wavelength $\lambda_M$ (Eq. \ref{EQ_Wien}) within the visual band limited by the receiver's filter $x$. With exception of pulsar class, referenced stellar systems  have fixed and unique peak wavelength that do not change significantly over traveling time. For example, our Sun spot cycle is about 11 years which alters the surface temperature and effects solar irradiance output by only $0.07\%$ \cite{REF_SolarCycle}.

Under Special Relativity, relativistic Doppler effect is caused by the relative motion between the referenced star and moving spaceship. As the spaceship approaches the star, the arrival light wavefront decreases in wavelength, a physics phenomenon called blueshift; and moving apart is redshift, reference to ``Fig.~\ref{VisibleLight} Visible Spectrum". Spaceship traveling velocity $v$ is expressed as Lorentz factor $\gamma$ and $\beta$ relative to the speed of light $c$ by their relationship:

\begin{equation}\label{EQ_LorentzFactor}
\gamma = \frac{1}{\sqrt{1-\beta^2}}
\end{equation}

\begin{equation}\label{EQ_LorentzBetaFactor}
\beta = \frac{v}{c}
\end{equation}

The laws of Relativistic Doppler effects for arriving light wavelength $\lambda$:

\begin{equation}\label{EQ_RelDoppler}
\Delta \lambda = \beta \cdot \lambda
\end{equation}

From polar coordinate of referenced star, we know the angle of approach $\theta$ and compute the relativistic Doppler shift as:

\begin{equation}\label{EQ_DopplerShift}
\Delta \lambda = \lambda \cdot cos(\theta) \sqrt{1-\frac{1}{\gamma^2}} 
\end{equation}

For error correction and navigation using relativistic Doppler shift \cite{REF_CelestialShift}, the equation can be inverted to correlate traveling speed to approaching angle with measured peak wavelength as:

\begin{equation}\label{EQ_InvertDopplerShift}
\theta = arccos\bigg(\frac{\Delta \lambda}{\lambda} \cdot (1-\frac{1}{\gamma^2})^{-\frac{1}{2}}\bigg)
\end{equation}

\begin{figure}[htbp]
\centerline{\includegraphics[width=3.5in]{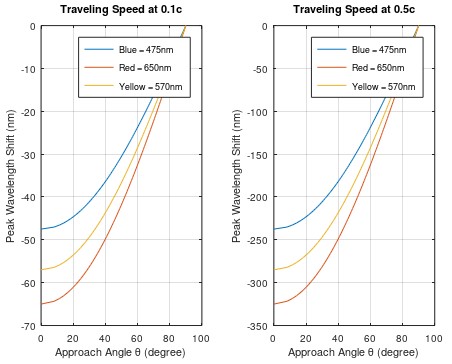}}
\caption{Peak Spectral Blueshift Plot as Function of $\theta$ Angle of Approach.}
\label{AppoarchAngle}
\end{figure}

Take a study case of three stars with their spectral peak wavelength at 475nm (blue), 575nm (yellow), and 650nm (red). We plot their peak wavelength blueshift as a function of approaching angle $\theta$, ``Fig.~\ref{AppoarchAngle}" for $\beta$ relative approaching speed at 0.1$c$ and 0.5$c$. The blueshift is non-linear, proportional to the traveling speed $\beta$ and inversely proportional at approaching angle $\theta$ with maximum blueshift occurring when the spaceship directly approaches the star.

Thus we demonstrate that we can improve travel vector correlation and navigational error correction to spaceship's GNC systems with only two unique referenced stars with different peak wavelength $\lambda_m$ and FOV angle $\theta$. The spaceship can effectively navigate in hyperspace using a few known celestial objects as references by exploiting the relativistic Doppler shift, proportional to the relative traveling speed and Lorentz factor $\beta$. Relativistic Doppler shift works for both forward and backward FOV of the spacecraft, exploiting both redshift and blueshift phenomenons as stars are receding and approaching traveling spacecraft.

\section{Navigation by Stellar Parallax}

\subsection{Solar Parallax}
\begin{figure}[htbp]
\centerline{\includegraphics[width=3in]{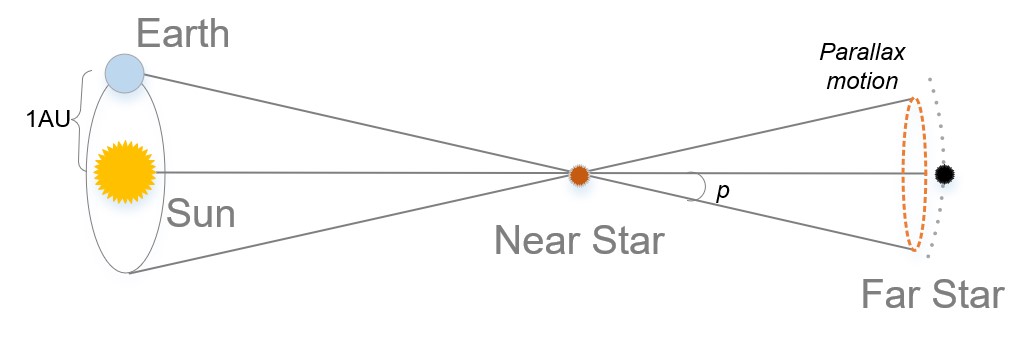}}
\caption{Solar Parallax.}
\label{SolarParallax}
\end{figure}

Solar parallax is a method for determining distance to nearby stars against static background using Earth-Sun known orbit; stellar parallax is the measuring of apparent shift of nearby stellar systems against the cosmic background as for determining the distance of the object using trigonometric parallax \cite{REF_Parallax}. From Earth, it takes half a year or half an Earth-Sun orbit to measure a nearby star against cosmic background, ``Fig.~\ref{SolarParallax}".

Solar parallax is useful to measure nearby star systems in terms of parallax angle in arc second (arcsec) using astronomical unit (AU), the norm distance between Earth and Sun. Parallax angle, $p$ (arcsec), increases as the distance to measuring object becomes closer to Earth.

\begin{equation}\label{EQ_SolarParallax}
d = \frac{1}{p}
\end{equation}

\subsection{Exploiting Stellar Parallax}
Stellar parallax is difficult because it requires a known orbital reference object; and it is only useful to measure distance to nearby stellar objects that are not in the stellar orbit with referencing background objects. Interstellar traveling spaceships would have difficulty finding nearby known reference orbiting objects in known star catalogs, especially when spaceships are exploring uncharted systems.

\begin{figure}[htbp]
\centerline{\includegraphics[width=2in]{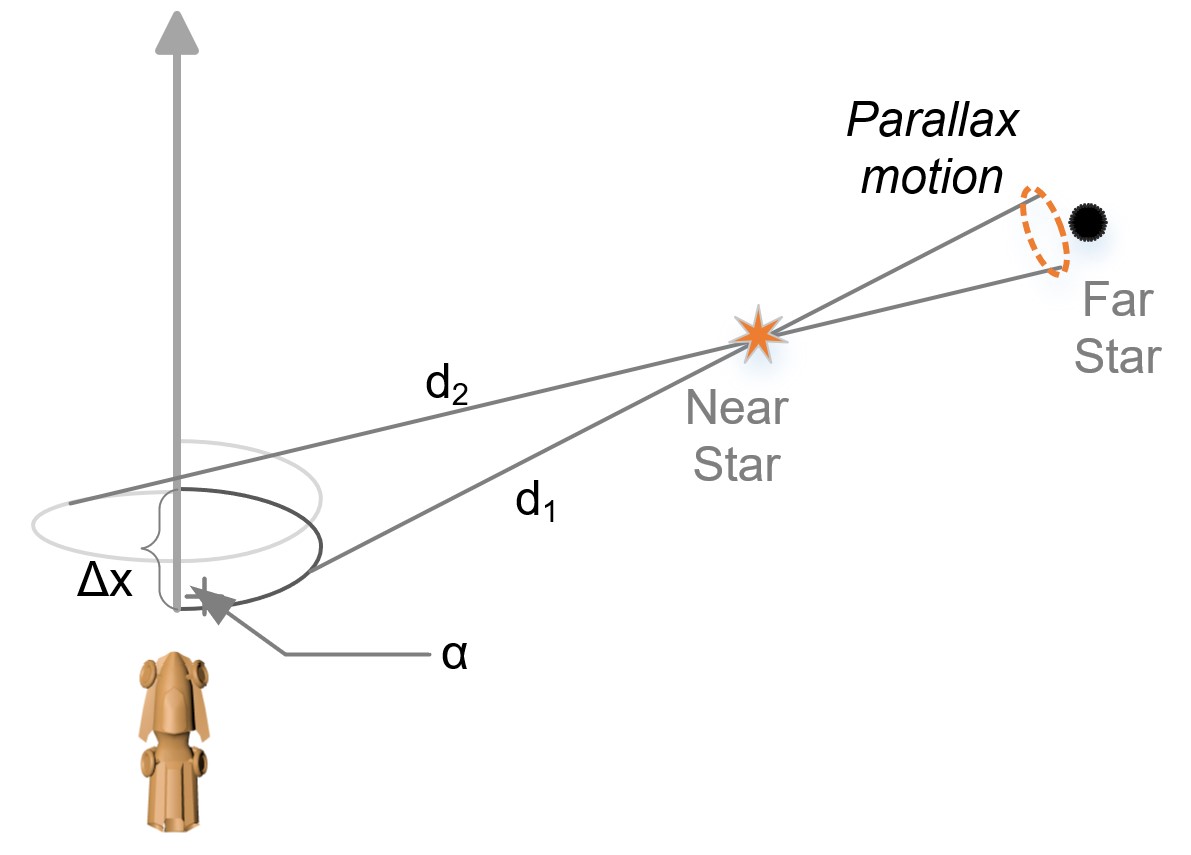}}
\caption{Travel Vector with $\alpha$ Spiral to Create Parallax.}
\label{Parallax}
\end{figure}

We propose a novel navigating method to apply trigonometric parallax to any stars within spaceship FOV. As spaceship rotates with angular rate $\omega$ inline to traveling vector, we already know which stellar objects are nearby by their relative viewing angle $\Delta \theta$ and relativistic Doppler spectral shift; the larger $\Delta \theta$, the smaller the spectral shift, and the closer stellar objects are to the traveling spacecraft. Therefore if we make a spacecraft traveling in a spiral with major angle $\alpha$ with an radius $r_{\alpha}$, it effectively allows this spiral trajectory to make parallax measurements to nearby stellar objects within the spacecraft FOV, performing both ranging and angular measurements, ``Fig.~\ref{Parallax}".

Exploiting relativistic Doppler shift, an interstellar traveling spaceship can measure the traveling angle to nearby referenced star. Adding major traveling spiral with $\alpha$ rotational rate over $\Delta x$ distance effectively create parallax for trilateration to determine spaceship coordinates. By exploiting both parallax and spectral relativistic shift \cite{REF_MultispectralParallax}, the spaceship can accurately compute ranging $d_s$ to this referenced star against far cosmic background, and accounting for slant range $d_1$ and $d_2$ from $\Delta x$. Our simulations demonstrate both major spiral rotation angle $\alpha$ and minor $\omega$ produced relativistic spectral shift $\Delta \lambda$ that correlated to measured $\Delta \theta$ and $\beta$, enabling parallax exploitation with large $\Delta \theta$ proportional to traveling speed $\beta$.

This method of spiral traveling to exploit stellar parallax only needs one nearby star. The parallax angle $p$ is proportional to the distance displacement $\Delta x$ and FOV angle $\theta$. Thus the faster the spacecraft travels, the better navigation and ranging from parallax. Most critically, the larger FOV angle $\theta$ is preferred for both Alcubierre's warp drive and hyperdrive with limited forward and backward FOV, ``Fig.~\ref{AlcubierreFOV}".

\section{Improving Ranging Using Variable Celestial X-ray Sources}
\begin{figure}[htbp]
\centerline{\includegraphics[width=2in]{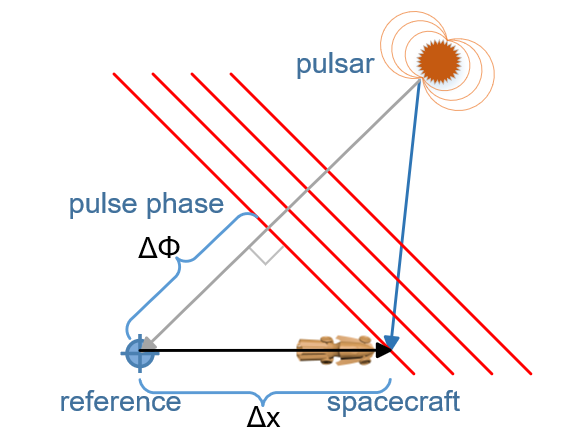}}
\caption{Pulsar ranging from a reference point and spacecraft position.}
\label{PulsarFigRanging}
\end{figure}

An existing method of using variable celestial x-ray sources such as pulsars can be used to improve ranging accuracy during hyperdrive \cite{REF_absolute}. This method requires known pulsars with predicted periodic signatures $\Phi$ (time and phase) and visible within the spacecraft's usable FOV ``Fig.~\ref{PulsarFigRanging}".

The primary benefit of using variable celestial sources is such that range determination achieves ``higher precision in the radial dimension than in the two transverse directions" \cite{REF_DeepSpace}. Hyperdrive limits usable FOV to a higher angle of approach $\theta$ relative to reference celestial objects. As such this method reduces dominant error propagation in the radial dimension due to lack of available celestial objects in FOV with small $\theta$ in transverse direction.

\begin{figure}[htbp]
\centerline{\includegraphics[width=2in]{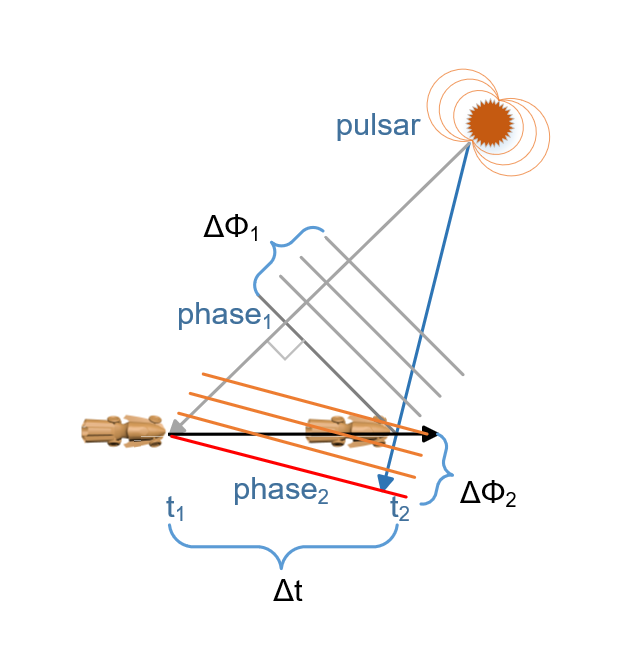}}
\caption{Pulsar ranging from TDOA with relativistic shift.}
\label{PulsarFigRedShift}
\end{figure}

At luminal and super-luminal  traveling speed, this exploit eliminates the need for a fixed known reference time source against as measurement to variable celestial source. As spacecraft traveled in hyperspace, relative distance and phase of the spacecraft over time are useful for Time Difference of Arrival method (TDOA) for ranging determination by exploiting Doppler shift $\Delta \Phi$, with spacecraft's onboard clock acted as reference timing source against pular's longer periodic signatures, ``Fig.~\ref{PulsarFigRedShift}". 

We can exploit geometric relationships of relative displacements in transverse $\Delta x$ and radial $\Delta d$, applying relativistic effect of spacecraft-pulsar TDOA and phase with $\Delta \theta$ angle to celestial objects. Relationship to navigation error or position dilution, is increasingly proportional to celestial distance $d$ but inversely to traveling speed $\beta$ and angle vector $\theta$, as well measuring time $\Delta t$ relative to pulsar's periodic $\Phi$. As such spacecraft's GNC benefits from precised ranging with correlated traveling speed $\beta$, reducing position dilution of measurement errors in both transverse direction $\Delta x$ by exploiting minor rotation $\omega$, and radial dimension with pulsar's ranging \cite{REF_absolute} correlated with $\Delta \theta$ to celestial objects.

Lastly, relativistic Doppler shift by Special Relativity apply equally to x-ray spectra of known pulsars. The measured relativistic Doppler shift of the same celestial source further reduces ranging error propagation for both range single and multiple differences \cite{REF_absolute}. Previous exploits using relativistic Doppler shift and parallax require the spacecraft to carry internal timing reference. Pulsar's known periodic phase carry inherent timing reference that spacecraft can lock-onto and improve geopositioning and correlate onboard timing clocks against these known timing reference sources that correlated to relativistic shift in phase and spectral peak. 

\section{2D FOV Simulation by First Order Approximation}

We present our simulations in Matlab by First Order approximation that accounted for relativistic Lorentz boost of a spaceship to a reference celestial and spectral Doppler shift of celestial object's luminous peak \cite{REF_Code}; we did not account for space expansion nor gravitational curvature of space (gravitational lensing). Our simulation codes and results are publicly available as part of Open Science Framework projects \cite{REF_OSF}.

From previous section of Eq. \ref{EQ_MagnitudeDistance}, a celestial object with relative 3D Euclidean coordinate in spaceship frame is projected onto 2D Cartesian FOV, ``Fig.~\ref{2DFOV}", by relationship

\begin{equation}\label{EQ_2DCartesian}
\begin{aligned}
azimuth = & d * cos (\theta) \\
altitude = & d * sin (\theta)
\end{aligned}
\end{equation}

The Cartesian coordinate pair (azimuth, altitude) or (x, y) can be converted to polar coordinates ($\rho$, $\phi$)

\begin{equation}\label{EQ_2DPolar}
\begin{aligned}
\rho = & \sqrt{(x^2 + y^2)} \\
\phi = & arctan(\frac{y}{x})
\end{aligned}
\end{equation}

Thus centric factor of celestial object in spaceship 2D FOV is closer to Cartesian origin coordinate (0,0) when ($\rho$, $\phi$) approaches zero. In first order, relativistic Doppler effect changes apparent magnitude $m$ with peak wavelength color representing  blueshift  when celestial object approaching, or redshift when receding.

\subsection{Minor Rotation Rate $\theta$ Projected onto 2D FOV}
\begin{figure}[htbp]
\centerline{\includegraphics[width=3.5in]{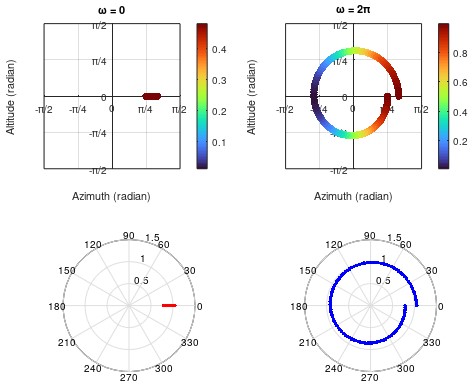}}
\caption{Minor Rotation $\omega=0$ and $\omega=2\pi$ rates.}
\label{minorRotation}
\end{figure}

Simulating for ``Fig.~\ref{StarFOVrotate}", we demonstrate by adding minor rotation $\omega$ around the traveling vector of spacecraft, displaying $\Delta \theta$ transit by celestial object for spaceship's $\Delta x$ displacement. ``Fig.~\ref{minorRotation}" is the simulation result with $\theta_1 = 45^{\circ} $ and $\theta_2 = 60^{\circ} $, with minor rotation $\omega$ rate at $0$ and $2\pi$ per $\Delta x$ linear distance by traveling speed $\beta=0.5c$, shown in both 2D Polar and Cartesian form; color bar of Cartesian plots is relative spectral Doppler shift of celestial spectral peak of 700mn. 

For eccentric expansion of star in 2D FOV, the spaceship approaching angle $\theta$ to celestial object is increasing while during concentric contraction, $\theta$ is decreasing until FOV = (0,0) when the celestial object is directly in the traveling path. The rate of eccentric and concentric is $\omega$ rate in radians.

\subsection{Major Spiral Rate $\alpha$ Projected onto 2D FOV}

\begin{figure}[htbp]
\centerline{\includegraphics[width=3.5in]{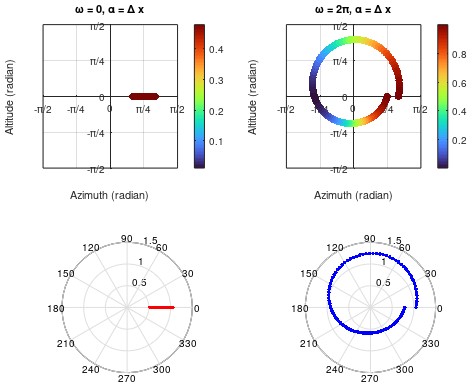}}
\caption{Major Rotation $\alpha$ and Minor $\omega$.}
\label{majorRotation}
\end{figure}

From previous simulation, we add a major spiral rotation $\alpha$ around traveling vector with diameter $\Delta x$ at $2\pi$ rate, ``Fig.~\ref{majorRotation}". This simulation demonstrates relationship of traveling spiral maximum distance $d_2$ to celestial object or ebb (receding), and minimum distance $d_1$ or flow (ascending) as celestial object transiting $2\pi$ rotation projected in 2D FOV in spaceship frame. 

\section{Conclusion}

We present our theoretical and simulation models in first order approximation with only three input factors: Lorentz $\gamma$, major $\alpha$, and minor $\omega$. Our models produce three outputs that are correlated to the eccentric and concentric of light bands, ebbing (receding) and flow (ascending) of celestial transits, and centric of celestial coordinate (azimuth and altitude). Finally our models show that both near-field and far-field celestial objects enhanced GNC's navigation correction and position dilution even when traveling speed approaches the speed of light. We demonstrate that our method of varying rotational rate $\omega$ and angle of approach $\alpha$ is superior to traveling at a straight line, especially when the Lorentz factor $\gamma$ is approaching $c$.

\section*{Acknowledgment}
Author acknowledges and is grateful for math and grammar corrections, and Matlab code development from physicist Timofey Golubev of Michigan State University during manuscript writing.


\begin{thebibliography}{00}
\bibitem{REF_hawking1973large} S. W. Hawking, and G. R. Ellis. ``The large scale structure of space-time". Cambridge University Press, 1973.
\bibitem{REF_EinsteinSingularities} J. Earman, and J. Eisenstaedt. ``Einstein and singularities". Studies In History and Philosophy of Modern Physics, vol. 30, p. 185-235, 1999.
\bibitem{REF_Wormsholes} M. Morris, K. Thorne, and U. Yurtsever. ``Wormholes, time machines, and the weak energy Condition". Phys. Rev. Lett., vol. 61, p. 1446-1449, 1988.
\bibitem{REF_Alcubierre} M. Alcubierre. ``The warp drive: hyper-fast travel within general relativity". Classical and Quantum Gravity, vol. 11, p. L73, 1994.
\bibitem{REF_NguyenKey} N. C. Nguyen. ``Key technological developments enabling human cosmic flight". International Astronautical Congress, IAC-22-A1.IP.28.x67262, 2022.
\bibitem{REF_NguyenNearField} N. C. Nguyen. ``Transforming near-field micro-gravity sources into far-field life support Systems". International Astronautical Congress, IAC-22-A1.IPB.11.x67265, 2022.
\bibitem{IMG_LightSpectrum} ``Visible spectrum of light.” Encyclopædia Britannica, https://www.britannica.com/science/color/The-visible-spectrum\#/media/1/126658/91330. Accessed 31 December 2022. 
\bibitem{IMG_WienLaw} ``Blackbody radiation.” Encyclopædia Britannica, https://www.britannica.com/science/Wiens-law\#/media/1/643338/127565. Accessed 31 December 2022.  
\bibitem{IMG_NickStrobelHR} ``Nick strobel's astronomy notes.” Nick Strobel, https://www.astronomynotes.com. Accessed 11 January 2023.
\bibitem{IMG_HRdiagram} ``Actual HR diagram based on Hipparcos data.” European Space Agency, https://sci.esa.int/web/education/-/35774-stellar-radiation-stellar-types?section=hertzsprung-russell-diagram. Accessed 11 January 2023.
\bibitem{REF_SolarCycle} C. D. Camp, and K. Tung. ``Surface warming by the solar cycle as revealed by the composite mean difference projection". Geophysical Research Letters, vol. 34, no. 14, 2007.
\bibitem{REF_CelestialShift} X. Chen, Z. Sun, Q. Huang, M. Liu and W. Zhang, "Hardware in-the-loop simulation of celestial angle and velocity measurement integrated navigation system," 2018 37th Chinese Control Conference, Wuhan, China, 2018, pp. 4821-4826, doi: 10.23919/ChiCC.2018.8483167.
\bibitem{REF_Parallax} A. Hirshfeld. ``Parallax: the race to measure the cosmos". W. H. Freeman, 2001.
\bibitem{REF_MultispectralParallax} X. Tong et al., "Attitude oscillation detection of the ZY-3 satellite by using multispectral parallax images," IEEE Transactions on Geoscience and Remote Sensing, vol. 53, no. 6, pp. 3522-3534, June 2015, doi: 10.1109/TGRS.2014.2379435.
 \bibitem{REF_absolute} S. I. Sheikh, , A. R. Golshan, and D. J. Pines. ``Absolute and relative position determination using variable celestial x-ray sources." 30th Annual AAS Guidance and Control Conference. Breckenridge, Colorado: American Astronautical Society, 2007.
\bibitem{REF_DeepSpace} P. S. Ray, S. I. Sheikh, P. H. Graven, M. T. Wolff, K. S. Wood, and K. C. Gendreau. ``Deep space navigation using celestial x-ray sources." Proceedings of the 2008 National Technical Meeting of The Institute of Navigation, 2008.
\bibitem{REF_Code} Hsss Science Code Repository, https://github.com/NghiHsss/Hsss-Science-Public/. Accessed 1 Febuary 2023.
\bibitem{REF_OSF} ``Hyperspace and interstellar travel project", Open Science Framework, https://osf.io/sx5rq/, doi: 10.17605/OSF.IO/SX5RQ.

\end{thebibliography}
\end{document}